\documentclass[12pt]{article}
\usepackage{graphicx}

\def\institute{Inter-University Institute for High Energies, Vrije Universiteit Brussel\\
Pleinlaan 2, 1050 Brussels, Belgium}

\def\Title#1{\begin{center} {\Large #1 } \end{center}}
\def\Author#1{\begin{center}{ \sc #1} \end{center}}
\def\Address#1{\begin{center}{ \it #1} \end{center}}

\newenvironment{Abstract}{\begin{quotation}  }{\end{quotation}}
\newenvironment{Presented}{\begin{quotation} \begin{center} 
             PRESENTED AT\end{center}\bigskip 
      \begin{center}\begin{large}}{\end{large}\end{center} \end{quotation}}

\def\beq{\begin{equation}}
\def\eeq#1{\label{#1}\end{equation}}
\def\eeqn{\end{equation}}

\def\beqa{\begin{eqnarray}}
\def\eeqa#1{\label{#1}\end{eqnarray}}
\def\eeqan{\end{eqnarray}}

\let\bar=\overbar

\def\Dslash{\not{\hbox{\kern-4pt $D$}}}
\def\dslash{\not{\hbox{\kern-2pt $\del$}}}

\def\msb{{\bar{\ssstyle M \kern -1pt S}}}

\begin{document}
\begin{titlepage}

\vfill
\Title{First observation of the $\mathrm{t\bar{t}H}$ process at CMS}
\vfill
\Author{Kirill Skovpen\\(on behalf of the CMS Collaboration)}
\Address{\institute}
\vfill
\begin{Abstract}
The top quark and the Higgs boson play a special role in the
fundamental interactions of the standard model. The observation of the
top quark pair production in association with the Higgs boson
establishes the first direct measurement of the tree-level coupling of the
Higgs boson to the top quark. The analysis of the data collected by
the CMS detector results in the overall observed significance of
5.2 standard deviations for this process.
\end{Abstract}
\vfill
\begin{Presented}
$11^\mathrm{th}$ International Workshop on Top Quark Physics\\
Bad Neuenahr, Germany, September 16--21, 2018
\end{Presented}
\vfill
\end{titlepage}
\def\thefootnote{\fnsymbol{footnote}}
\setcounter{footnote}{0}

\section{Introduction}

The top quark and the Higgs boson are the two heaviest elementary
particles of the standard model (SM). The top quark, the heaviest of all elementary particles, decays before any
hadronization takes place and provides an access to a direct study of
its intrinsic properties in experiment. The second heaviest elementary
particle that was recently discovered, the Higgs boson, gives mass to all SM particles through the
omnipresent Higgs field~\cite{higgs1,higgs2,higgs3}. It is of particular interest to probe the strength of the interaction between
these two heavyweight elementary objects: due to the large mass of the
top quark and the fact that the strength of Yukawa interaction is proportional to the
mass of the quark that interacts with the Higgs field, the top quark
is expected to have a large Yukawa coupling, $y_{t} \simeq 1$.

The precise determination of $y_{t}$ represents a direct test of 
the SM predictions, as well as a probe of new physics. Moreover, the
properties of this fundamental interaction are potentially connected to
the vacuum stability of the Higgs potential. High precision
measurements of the masses of the top quark and the Higgs boson, as
well as the strength of the top-Higgs interaction, can shed some light
on the stability state of our universe at the Planck
scale~\cite{vacuum1,vacuum2}.

The direct measurement of $y_{t}$ is possible from the study of the
process of associated production of the Higgs boson with a pair of
top quarks ($\mathrm{t\bar{t}H}$), as well as in the associated production
of the single top quark with the Higgs boson ($\mathrm{tHq}$). However,
the $\mathrm{tHq}$ process is strongly suppressed due to destructive
interference present in the production diagrams.

\section{Search channels}

The $\mathrm{t\bar{t}H}$ production is a rare process with the predicted
cross section of $\simeq 0.5$~pb at 13 TeV in the center-of-mass of
proton-proton collisions at the LHC.
There are several decays of the Higgs boson that define three main
search channels considered in analysis of recorded data by 
the CMS detector~\cite{det}. The channel with $H \rightarrow bb$ benefits
from the large branching fraction of this decay process but suffers from high
background rates. The $H \rightarrow \gamma\gamma$ channels provides a 
relatively clean experimental signature but is also attributed to rather
small branching ratio of this decay channel. Finally, the $H
\rightarrow WW^{*},ZZ^{*},\tau\tau$, or multilepton channel, represents
a compromise between the modest background rates and
reasonably high rates of signal events. In the combined analysis of the 8 TeV
data collected by ATLAS and CMS detectors, the observed (expected)
significance of 4.4 (2.0) standard deviations ($\sigma$) was obtained
for the $\mathrm{t\bar{t}H}$ process~\cite{ttHRun1}.

The search for $\mathrm{t\bar{t}H}$ is performed for $H
\rightarrow bb$ decays with at least one lepton present in the final
state~\cite{ttHbbLep}. With the requirement of at least three b-tagged jets to be
identified in association with one or two leptons, the
dominant background becomes top quark pair production with additional
jets ($\mathrm{t\bar{t}}$+jets). In the first step of the event
categorisation procedure the events are classified based on the number
of b-tagged jets and leptons to define a multivariate analysis (MVA), deep neural network (DNN) and
matrix-element method (MEM) discriminants, with the final choice of the method driven
by the obtained highest expected sensitivity for the signal process.
In the single lepton case the multi-classification procedure is used
to define several physics process categories corresponding to the
signal and $\mathrm{t\bar{t}}$+jets background processes, split by
the flavour of additional jets. The DNN
discriminant is used as a final variable to perform a maximum
likelihood fit. In the dilepton selection, the BDT and MEM discriminants are used.
The simultaneous combined likelihood fit is performed over all
considered event categories. The final observed (expected)
significance is 1.6 (2.2)~$\sigma$.

The search for $\mathrm{t\bar{t}H}$ in $H
\rightarrow bb$ decays is also done in the fully hadronic
channel with considering only hadronic decays of the top
quarks~\cite{ttHbbHad}. Even though the combinatorial background
represents an important challenge in this
analysis, the full event reconstruction is possible.
Events with reconstructed leptons are vetoed, and the dominant background arises from
the QCD multijet production. This background is estimated from the
dedicated control regions in data corresponding to the low b tag
multiplicity. Six exclusive categories are defined based on the
number of b and non-b-tagged jets. The final signal contribution is
extracted from a maximum likelihood fit to the MEM
discriminant which is based on the probabilities of an event
to correspond to either the $\mathrm{t\bar{t}H}$ or
$\mathrm{t\bar{t}}$+$b\bar{b}$ hypothesis for the
production process. the best fit signal strength is compatible with the
background-only hypothesis and therefore only the
observed (expected) upper limit at 95\% confidence level of 3.8 (3.1)
can set on the signal strength.

One of the search channels where one can aim at observing a
reconstructed mass peak corresponding to the Higgs boson production is
the channel with Higgs decaying to a pair of photons~\cite{ttHgg}.
The background processes consist of irreducible prompt diphoton
production and reducible backgrounds associated with $\gamma$+jet
and di-jet events, in which jets are misidentified as isolated photons. 
Identification of primary vertex with two photons is performed with 
boosted decision trees (BDT) and has a strong impact on the
reconstructed di-photon invariant mass resolution. The BDT uses several 
observables including the information on the tracks recoiling against the di-photon system. 
The background model is defined from fitting data to extract the signal from a
maximum likelihood fit which is performed on the di-photon mass
distribution. Evidence for the $\mathrm{t\bar{t}H}$ process is obtained
with an observed (expected) significance of 3.3 (1.5)~$\sigma$.

The analysis of the multilepton final states is sensitive to
$H \rightarrow WW^{*},ZZ^{*},\tau\tau$ decays~\cite{ttHML}. 
Six exclusive categories are defined based on the number of
reconstructed electrons, muons and hadronic tau decays, starting from
the single lepton selection and going up to the four-lepton requirement. 
The inclusion of hadronic decays of taus in the analysis allows to
significantly enhance the signal sensitivity to $H \rightarrow \tau\tau$.
The dominant background is associated with the $\mathrm{t\bar{t}}$+Z
and $\mathrm{t\bar{t}}$+W processes, as well as misidentified leptons.
The lepton misidentification and charge flip probabilities are estimated with
data-driven approaches. A dedicated MVA is used for the optimized lepton selection and is based
on several variables associated with the reconstructed lepton and the properties 
of the closest reconstructed jet.
The MEM is additionally used in the event categories with three
leptons, and all channels exploit BDT as the final discriminating variable,
except for the four lepton selection where the cut-and-count analysis is
performed. The extracted signal contribution from the maximum
likelihood fit corresponds to the observed (expected) significance of
3.2 (2.8)~$\sigma$. This result also leads to the observed evidence for the
$\mathrm{t\bar{t}H}$ process.

\section{Combination}

\begin{figure}[htb]
\centering
\includegraphics[height=3.5in]{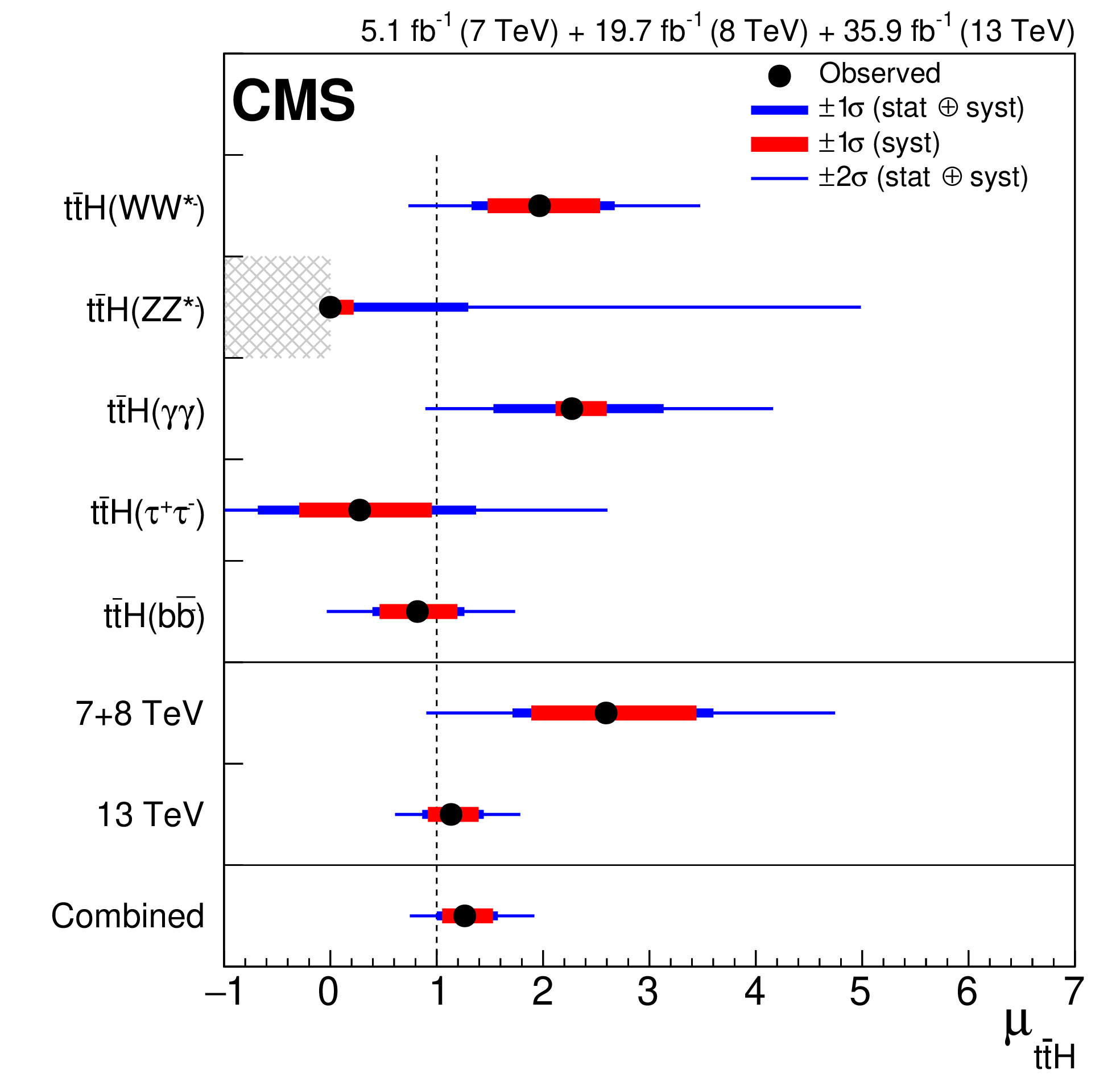}
\caption{Best fit value for the $\mathrm{t\bar{t}H}$ signal strength
modifier~\cite{ttHCMS}. The SM prediction is shown as a dashed vertical line.}
\label{fig:fig1}
\end{figure}

\begin{figure}[htb]
\centering
\includegraphics[height=3.5in]{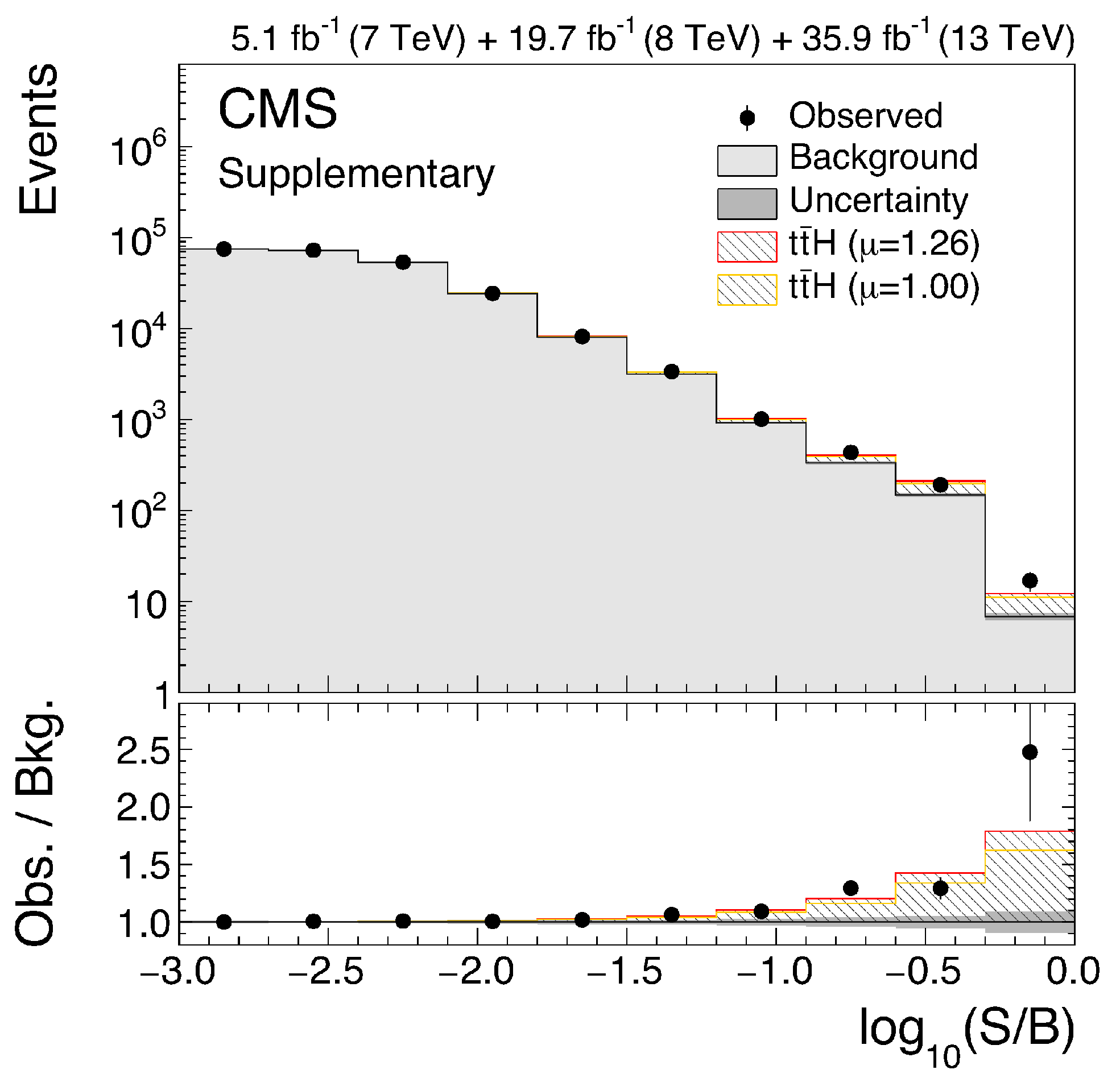}
\caption{Number of events as a function of $\log{S/B}$, where $S$ and $B$
are the signal and background yields, respectively~\cite{ttHCMS}.}
\label{fig:fig3}
\end{figure}

The results obtained in each of the search channels are combined in a
maximum likelihood fit. The fitted
signal strength values for each channel, as well as for the
combination, are represented in Fig.~\ref{fig:fig1}.
The observed and predicted event yields in all the bins of the 
final discriminants are ordered by the pre-fit expected signal-to-background ratio. 
The result distribution is shown in Fig.~\ref{fig:fig3}. The overall observed (expected)
significance is 5.2 (4.2)~$\sigma$ with the measured production rate consistent 
with the SM prediction within one standard
deviation. This result represents the first experimental observation of the
$\mathrm{t\bar{t}H}$ process.

\section{Search for the $\mathrm{tHq}$ process}

The single top associated production with the Higgs boson is an
important process that is sensitive not only to the absolute value of
$y_{t}$, but also to the sign of this parameter. However, this is a
very rare process with the predicted cross
section that is almost one order of magnitude smaller than that
of $\mathrm{t\bar{t}H}$. The search for $\mathrm{tHq}$ process was done with
8 and 13 TeV data in the similar final states as considered in the
$\mathrm{t\bar{t}H}$ analysis ($H \rightarrow
bb,\gamma\gamma,WW^{*},ZZ^{*},\tau\tau$), and led to stringent
contraints set on the anomalous values of
$y_{t}$~\cite{tHq8TeV,tHq13TeV}.
A positive value of the modifier of the top-Higgs coupling is favoured
by about 1.5 standard deviations with excluding values outside the
ranges of [-0.9, -0.5] and [1.0, 2.1] at 95\% C.L. The standard model-like 
Higgs couplings to vector bosons are assumed.

\section{Conclusion}

The observation of the $\mathrm{t\bar{t}H}$ process represents an
important milestone in our study of the top-Higgs interactions, and it is the first step towards our
better understanding of the properties of this fundamental interaction. The analysis of
the full data set collected at 13 TeV, as well as the future
experiments at the LHC, will allow us to significantly improve the
precision of the current measurements of the $\mathrm{t\bar{t}H}$ production
cross section and to study differential distributions of the
particles involved in this process with a potential probe of new physics.

\end{document}